\documentclass[10pt]{article}
\usepackage{euscript,amssymb,epsfig}
\usepackage{amsfonts,latexsym}
\usepackage{euscript,amssymb}
\usepackage{epsfig}
\usepackage{amsmath}
\usepackage{graphicx}
\newcommand{\etal}{{\it et al }}

\catcode`\@=11

\def\vereq#1#2{\lower3pt\vbox{\baselineskip1.5pt \lineskip1.5pt
\ialign{$\m@th#1\hfill##\hfil$\crcr#2\crcr\sim\crcr}}}

\def\Let@{\relax\iffalse{\fi\let\\=\cr\iffalse}\fi}
\def\vspace@{\def\vspace##1{\crcr\noalign{\vskip##1\relax}}}
\def\multilimits@{\bgroup\vspace@\Let@
 \baselineskip\fontdimen10 \scriptfont\tw@
 \advance\baselineskip\fontdimen12 \scriptfont\tw@
 \lineskip\thr@@\fontdimen8 \scriptfont\thr@@
 \lineskiplimit\lineskip
 \vbox\bgroup\ialign\bgroup\hfil$\m@th\scriptstyle{##}$\hfil\crcr}
\def\Sb{_\multilimits@}
\def\endSb{\crcr\egroup\egroup\egroup}
\def\Sp{^\multilimits@}

\def\RR{{\mathbb R}}

\catcode`\@=11

\renewcommand\footnoterule{%
  \kern-3\p@
  \hrule\@width.4\columnwidth
  \kern2.6\p@}
\renewcommand\@makefntext[1]{%
    \parindent 1em\noindent
    \hb@xt@1.8em{\hss$^{\@thefnmark}$)}\hspace{2pt}%
    \footnotesize\rmfamily#1}  %\@makefnmark}#1}
\def\@makefnmark{\hspace{.5pt}\hbox{$^{\@thefnmark}$%
\hspace{-1pt})}} 
\newcommand{\be}[1]{\begin{equation}\label{#1}}
\newcommand{\ee}{\end{equation}}
\newcommand{\ba}[1]{\begin{eqnarray}\label{#1}}
\newcommand{\ea}{\end{eqnarray}}
\newcommand{\rf}[1]{(\ref{#1})}
\newcommand{\nn}{\nonumber}

\newcommand{\sign}{ \mbox{\rm sign}\,}

\def\RR{\mathbb{R}}

\begin{document}

\title{AdS non-linear curvature-squared and curvature-quartic
multidimensional (D=8) gravitational models with stabilized extra
dimensions}

\author{Tamerlan Saidov$^a$\footnote{e-mail:
tamerlan-saidov@yandex.ru} \
 and Alexander Zhuk$^{a,b}$ \footnote{e-mail: zhuk@paco.net}\\[2ex] $^a$
Department of Theoretical Physics, Odessa National University,\\
2 Dvoryanskaya St.,
Odessa 65026, Ukraine \\
\\$^b$ Department of Advanced Mathematics, \\Odessa National Academy of Telecommunication,\\
1 Kuznechnaya St., Odessa 65069, Ukraine }

%\author{Tamerlan Saidov}\email{tamerlan-saidov@yandex.ru} \
%\author{Alexander Zhuk} \email{zhuk@paco.net}\\[2ex]
%\altaffiliation[Also at ]{Department of Advanced Mathematics,
%Odessa National Academy of Telecommunication, 1 Kuznechnaya St.,
%Odessa 65069, Ukraine}
%\affiliation{Department of Theoretical Physics, Odessa National University,\\
%2 Dvoryanskaya St.,
%Odessa 65026, Ukraine} \\

\date{}

\maketitle

%{\center{accelerate.tex}}

\begin{abstract}We investigate $D$-dimensional gravitational model with
curvature-quadratic and curvature-quartic correction terms:
$R+R^2+R^4$. It is assumed that the corresponding higher
dimensional spacetime manifold undergos a spontaneous
compactification to a manifold with warped product structure.
Special attention is paid to the stability of the
extra-dimensional factor space for a model with critical dimension
$D=8$. It is shown that for certain parameter regions the model
allows for a freezing stabilization of this space. The effective
four-dimensional cosmological constant is negative and the
external four-dimensional spacetime is asymptotically AdS.
\end{abstract}

%%%%%%%%%%%%%%%%%%%%%%%%%%%%%%%%%%%%%%%%%%%%%%%%%%%%%%%%%%%%%%%%%

\section{\label{sec:1}Introduction}

\setcounter{equation}{0}

Multidimensionality of our Universe is one of the most intriguing
assumption in modern physics. It follows from theories which unify
different fundamental interactions with gravity, such as M or
string theory \cite{GSW}, and which have their most consistent
formulation in spacetimes with more than four dimensions. Thus,
multidimensional cosmological models have received a great deal of
attention over the last years.

Stabilization of additional dimensions near their present day
values (dilaton/geometrical moduli stabilization) is one of the
main problems for any multidimensional theory because a dynamical
behavior of the internal spaces results in a variation of the
fundamental physical constants. Observations show that internal
spaces should be static or nearly static at least from the time of
recombination (in some papers arguments are given in favor of the
assumption that variations of the fundamental constants are absent
from the time of primordial nucleosynthesis \cite{Kolb}). In other
words, from this time the compactification scale of the internal
space should either be stabilized and trapped at the minimum of
some effective potential, or it should be slowly varying (similar
to the slowly varying cosmological constant in the quintessence
scenario). In both cases, small fluctuations over stabilized or
slowly varying compactification scales (conformal
scales/geometrical moduli) are possible.

Stabilization of extra dimensions (moduli stabilization) in models
with large extra dimensions (ADD-type models) has been considered
in a number of papers (see e.g., Refs. \cite{sub-mill2}-\cite{PS}
%\cite{sub-mill2,d2,sub-mill3,CGHW,Geddes,demir,NSST,PS}
). In the corresponding approaches, a product topology of the
$(4+D^{\prime })-$dimensional bulk spacetime was constructed from
Einstein spaces with scale (warp) factors depending only on the
coordinates of the external $4-$dimensional component. As a
consequence, the conformal excitations had the form of massive
scalar fields living in the external spacetime. Within the
framework of multidimensional cosmological models (MCM) such
excitations were investigated in
\cite{GZ(PRD1997)}-\cite{GZ(PRD2000)}
%\cite{GZ1,GZ,GZ(PRD2)}
where they were called gravitational excitons. Later, since the
ADD
%sub-millimeter weak-scale
compactification approach these geometrical moduli excitations are
known as radions \cite{sub-mill2,sub-mill3}.

Most of the aforementioned  papers are devoted to the
stabilization of large extra dimensions in theories with a linear
multidimensional gravitational action. String theory suggests that
the usual linear Einstein-Hilbert action should be extended with
higher order nonlinear curvature terms. In our papers
\cite{GMZ(PRDa)}-\cite{GZBR} we considered a simplified model with
multidimensional Lagrangian of the form $L = f(R)$, where $f(R)$
is an arbitrary smooth function of the scalar curvature. Without
connection to stabilization of the extra-dimensions, such models
($4-$dimensional as well as multidimensional ones) were considered
e.g. in Refs. \cite{Kerner}-\cite{EKOY}.
%\cite{Kerner,Maeda,EKOY}.
There, it was shown that the nonlinear models are equivalent to
models with linear gravitational action plus a minimally coupled
scalar field with self-interaction potential. Similar approach was
elaborated in Refs. \cite{NO} where the main attention was paid to
a possibility of the late time acceleration of the Universe due to
the nonlinearity of the model. In our papers
\cite{GMZ(PRDa)}-\cite{GZBR}, we advanced the equivalence between
the nonlinear models and the linear ones with a minimally coupled
scalar field towards investigating the stabilization problem for
extra dimensions\footnote{A different approach to the problem of
the extra dimension stabilization was proposed in Ref. \cite{BR}.
This method can be applied to Lagrangians containing high-order
curvature invariants. In the case of the models with Lagrangians
$L = f(R)$ both of these approaches result in the same
conclusions.}. Particular attention was paid to models where usual
linear curvature term $R$ was supplemented with either $R^2$ or
$R^4$ or $R^{-1}$ nonlinear terms. All of these models can be
investigated analytically. It was shown that for certain parameter
ranges, the extra dimensions are stabilized. In the present paper
we extend this consideration to a model with Lagrangian of the
type $f(R)=R+R^2+R^4$. Such simple generalization enriches
considerably the qualitative behavior of the model because it
results in either one-branch or three-branch models. For each of
these branches the stability analysis should be performed
separately. Our paper is the first one among two papers devoted to
this model. Here, we investigate the one-branch model postponing
the three-branch model for our forthcoming paper.  This model can
be solved analytically. However, this analysis is very cumbersome
for an arbitrary number of dimensions $D$. So, we restrict our
consideration to a critical dimension $D=8$. Critical dimension is
defined by doubled degree of the scalar curvature polynomial
$f(R)$, i.e. in our case $D=2\times 4$ (see \cite{GZBR} and
Appendix A below). We show that for certain parameter regions the
model allows for a freezing stabilization of the internal space.
Here, the stabilization takes place in negative minimum of the
effective potential. Thus the effective four-dimensional
cosmological constant is also negative and the homogeneous and
isotropic external four-dimensional spacetime is asymptotically
AdS.

The paper is structured as follows. In section \ref{setup}, we
present a brief technical outline of the transformation from a
non-Einsteinian purely gravitational model with general scalar
curvature nonlinearity of the type $f(\bar R)$ to an equivalent
curvature-linear model with additional nonlinearity carrying
scalar field. Afterwards, we derive criteria which ensure the
existence of at least one minimum for the effective potential of
the internal space scale factors (volume moduli). These criteria
are then used in Sec. \ref{model} to obtain the regions in
parameter space which allow for a freezing stabilization of the
scale factors in the model with scalar curvature nonlinearities of
the type $f(R)=R+R^2+R^4$. A brief discussion of the obtained
results is presented in the concluding Sec. \ref{conclu}. In
appendix \ref{app1} it is shown that $D = 2\times
\mbox{deg}_{\bar{R}}(f)$ is the critical dimension of the
considered models.  Finally, graphical visualization of the
effective potential with a global minimum is given in appendix
\ref{app2}.

%%%%%%%%%%%%%%%%%%%%%%%%%%%%%%%%%%%%%%%%%%%%%%%%%%%%%%%%%%%%%%%%%%%%

\section{\label{setup}General setup}
\setcounter{equation}{0}

We consider a $D= (4+D^{\prime})-$dimensional nonlinear pure
gravitational theory with action functional
%%%%%%
\be{1.1} S = \frac {1}{2\kappa^2_D}\int_M d^Dx \sqrt{|\bar g|}
f(\bar R)\; , \ee
%%%%%%
where $f(\bar R)$ is an arbitrary smooth function with mass
dimension $\mathcal{O}(m^2)$ \ ($m$ has the unit of mass) of a
scalar curvature $\bar R = R[\bar g]$ constructed from the
$D-$dimensional metric $\bar g_{ab}\; (a,b = 1,\ldots,D)$.
$D^{\prime}$ is the number of extra dimensions and $\kappa^2_D $
denotes the $D-$dimensional gravitational constant which is
 connected with the fundamental mass scale
$M_{*(4+D^{\prime})}$ and the surface area $S_{D-1}=2\pi
^{(D-1)/2}/\Gamma [(D-1)/2]$ of a unit sphere in $D-1$ dimensions
by the relation \cite{add1,GZ-mg10}
%%%%
\be{1.2}  \kappa^2_D = 2S_{D-1} /
M_{*(4+D^{\prime})}^{2+D^{\prime}}.\ee
%%%%
Before we endow the metric of the  pure gravity theory \rf{1.1}
with explicit structure, we recall that this $\bar R-$nonlinear
theory is equivalent to a theory which is linear in another scalar
curvature $R$ but which contains an additional self-interacting
scalar field. According to standard techniques
\cite{Kerner,Maeda}, the corresponding $R-$linear theory has the
action functional:
%%%%%%%%%%%%%%%
\be{1.6} S = \frac{1}{2\kappa^2_D} \int_M d^D x \sqrt{|g|} \left[
R[g] - g^{ab} \phi_{,a} \phi_{,b} - 2 U(\phi )\right]\; , \ee
%%%%%%
where
%%%%
\be{1.7} f'(\bar R) = \frac {df}{d \bar R} := e^{A \phi} > 0\;
,\quad A := \sqrt{\frac{D-2}{D-1}}\;  ,\ee
%%%%%
and where the self-interaction potential $U(\phi )$ of the scalar
field $\phi$ is given by
%%%%%
\ba{1.8-1} U(\phi ) &=&
 \frac12 \left(f'\right)^{-D/(D-2)} \left[\; \bar R f' - f\right]\;
 ,\label{1.8a}\\
&=& \frac12 e^{- B \phi} \left[\; \bar R (\phi )e^{A \phi } -
f\left( \bar R (\phi )\right) \right]\; , \quad B := \frac
{D}{\sqrt{(D-2)(D-1)}}\, .\label{1.8} \ea
%%%%%
 This scalar field
$\phi$ carries the nonlinearity degrees of freedom in $\bar R$ of
the original theory, and for brevity we call it the nonlinearity
field.The metrics $g_{ab}$, $\bar g_{ab}$   of the two theories
\rf{1.1} and \rf{1.6} are conformally connected by the relations
%%%%
\be{1.9} g_{ab} = \Omega^2 \bar g_{ab} = \left[ f'(\bar
R)\right]^{2/(D-2)}\bar g_{ab}\;.  \ee
%%%%

As next, we assume that the D-dimensional bulk space-time $M$
undergoes a spontaneous compactification
%\footnote{For a discussion
%of possible decompactification scenarios we refer to the recent
%work \cite{decomp-1}.}
to a warped product manifold
%%%%%
\be{1.18} M = M_0 \times M_1 \times \ldots \times M_n \ee
%%%%%
with  metric
%%%%%
\be{1.19} \bar g=\bar g_{ab}(X)dX^a\otimes dX^b=\bar
g^{(0)}+\sum_{i=1}^ne^{2\bar {\beta} ^i(x)}g^{(i)}\; . \ee
%%%%%
The coordinates on the $(D_0=d_0+1)-$dimensional manifold $M_0 $
(usually interpreted as our observable $(D_0=4)-$dimensional
Universe) are denoted by $x$ and the corresponding metric by
%%%%%
\be{1.20} \bar g^{(0)}=\bar g_{\mu \nu }^{(0)}(x)dx^\mu \otimes
dx^\nu\; . \ee
%%%%%
For simplicity, we choose the internal factor manifolds $M_i$ as
$d_i-$dimensional Einstein spaces with metrics
$g^{(i)}=g^{(i)}_{m_in_i}(y_i)dy_i^{m_i}\otimes dy_i^{n_i},$ so
that the relations
%%%%%
\be{1.21} R_{m_in_i}\left[ g^{(i)}\right] =\lambda
^ig_{m_in_i}^{(i)},\qquad m_i,n_i=1,\ldots ,d_i \ee
%%%%%
and
%%%%%
\be{1.22} R\left[ g^{(i)}\right] =\lambda ^id_i\equiv R_i
%\sim k r_i^{-2}
\;  \ee hold.
%%%%%
The specific metric ansatz \rf{1.19} leads to a scalar curvature
$\bar R$ which depends only on the coordinates $x$ of the external
space: $\bar R[\bar g] = \bar R(x)$. Correspondingly, also the
nonlinearity field $\phi$ depends on $x$ only: $\phi = \phi (x)$.

Passing from the $\bar R-$nonlinear theory \rf{1.1} to the
equivalent $R-$linear theory \rf{1.6} the metric \rf{1.19}
undergoes the conformal transformation $\bar g \mapsto g$ [see
relation \rf{1.9}]
%%%%%
\be{1.23} g = \Omega^2 \bar g = \left( e^{A \phi
}\right)^{2/(D-2)} \bar g\: := g^{(0)}+\sum_{i=1}^ne^{2
\beta^i(x)}g^{(i)}\;  \ee
%%%%%
with
%%%%
\be{1.24} g^{(0)}_{\mu \nu} := \left( e^{A \phi}\right)^{2/(D-2)}
\bar g^{(0)}_{\mu \nu}\; , \quad
 \beta^i := \bar {\beta} ^i + \frac{A}{D-2} \phi\; . \ee
%%%%%

The main subject of our subsequent considerations will be the
stabilization of the internal space components. A strong argument
in favor of stabilized or almost stabilized internal space scale
factors $\bar {\beta} ^i(x)$, at the present evolution stage of
the Universe, is given by the intimate relation between variations
of these scale factors and those of the fine-structure constant
$\alpha$ \cite{GSZ}. The strong restrictions on
$\alpha-$variations in the currently observable part of the
Universe \cite{alpha-var} imply a correspondingly strong
restriction on these scale factor variations \cite{GSZ}. For this
reason, we will concentrate below on the derivation of criteria
which will ensure a freezing stabilization of the scale factors.
Extending earlier discussions of models with $\bar{R}^2$ , $\bar
{R}^4$ and $\bar{R}^{-1}$ scalar curvature nonlinearities
\cite{GMZ(PRDa)}-\cite{GZBR} we will investigate here models of
the nonlinearity type $\bar{R}^{2}+\bar{R}^4$.

In Ref. \cite{GZ(PRD2000)} it was shown that for models with a
warped product structure \rf{1.19} of the bulk spacetime $M$ and a
minimally coupled scalar field living on this spacetime, the
stabilization of the internal space components requires a
simultaneous freezing of the scalar field. Here we expect a
similar situation with simultaneous freezing stabilization of the
scale factors ${\beta} ^i(x)$ and the nonlinearity field
$\phi(x)$. According to \rf{1.24}, this will also imply a
stabilization of the scale factors $\bar {\beta} ^i(x)$ of the
original nonlinear model.

In general, the model will allow for several stable scale factor
configurations (minima in the landscape over the space of volume
moduli). We choose one of them\footnote{Although the toy model
ansatz \rf{1.1} is highly oversimplified and far from a realistic
model, we can roughly think of the chosen minimum, e.g., as that
one which we expect to correspond to current evolution stage of
our observable Universe.}, denote the corresponding scale factors
as $\beta^i_0$, and work further on with the deviations
%%%%%%%%
\be{1.25} \hat \beta^i (x)= \beta^{i}(x) - \beta^{i}_0\ee
%%%%%%%%
as the dynamical fields. After dimensional reduction of the action
functional \rf{1.6} we pass from the intermediate Brans-Dicke
frame to the Einstein frame via a conformal transformation
\begin{equation}
\label{1.26}g_{\mu \nu }^{(0)}=\hat{\Omega} ^2\hat g_{\mu \nu
}^{(0)} =\left( \prod^{n}_{i=1} e^{d_i\hat \beta ^i}
\right)^{-2/(D_0-2)} \hat g_{\mu \nu }^{(0)}\,
\end{equation}
with respect to the scale factor deviations $\hat \beta^i (x)$
\cite{GMZ(PRDa)}-\cite{GZBR},\cite{GZ-mg10}. As result we arrive
at the following action
%%%%%
\begin{equation}
\label{1.27} S = \frac 1{2\kappa
_{D_0}^2}\int\limits_{M_0}d^{D_0}x\sqrt{|\hat g^{(0)}|}\left\{
\hat R\left[ \hat g^{(0)}\right] -\bar G_{ij}\hat g^{(0)\mu \nu
}\partial _\mu\hat \beta ^i\,\partial _\nu \hat \beta ^j - \hat
g^{(0)\mu \nu }\partial _\mu\phi \,\partial _\nu\phi
-2U_{eff}\right\} \, ,
\end{equation}
%%%%%
which contains the scale factor offsets $\beta^i_0$ through the
total internal space volume
%%%%%%
\be{1.29} V_{D'} \equiv V_I\times v_0\equiv
\prod^{n}_{i=1}\int_{M_i}d^{d_i}y \sqrt{|g^{(i)}|}\times
\prod^{n}_{i=1}e^{d_i\beta^{i}_0}\ee
%%%%%%
in the definition of the effective gravitational constant $\kappa
_{D_0}^2$ of the dimensionally reduced theory
%%%%
\be{1.28}\kappa _{(D_0=4)}^2=\kappa_D ^2/V_{D'}= 8\pi/M^2_{4}\quad
\Longrightarrow \quad M_{4}^2 = \frac{4\pi}{S_{D-1}}V_{D'}
M_{*(4+D^{\prime})}^{2+D^{\prime}}.\ee
%%%%
Obviously, at the present evolution stage of the Universe, the
internal space components should have a total volume which would
yield a four-dimensional mass scale of order of the Planck mass
$M_{(4)} = M_{Pl}$. The tensor components of the midisuperspace
metric (target space metric on $\RR _T^n$) reads: $\bar
G_{ij}=d_{i}\delta_{ij}+d_{i}d_{j}/(D_{0}-2)$, where $
i,j=(1,\ldots ,n)$, see \cite{IMZ,RZ}. The effective potential has
the explicit form
%%%%%%%
\begin{equation}
\label{1.31}U_{eff} ( \hat \beta , \phi ) ={\left(
\prod_{i=1}^ne^{d_i\hat \beta ^i}\right) }^{-\frac 2{D_0-2}}\left[
-\frac 12\sum_{i=1}^n\hat R_{i}e^{-2\hat \beta ^i}+ U(\phi )
\right] \, ,
\end{equation}
%%%%%%%
where we abbreviated
%%%%%
\be{1.31a} \hat R_{i} := R_i
\exp{(-2\beta^{i}_{0})}. \ee
%%%%

A freezing stabilization of the internal spaces will be achieved
if the effective potential has at least one minimum with respect
to the fields $\hat \beta^i(x)$. Assuming, without loss of
generality, that one of the  minima  is located at $\beta^{i} =
\beta^{i}_{0} \Rightarrow \hat \beta^{i} = 0$, we get the extremum
condition:
%%%%%%
\be{1.32} \left.\frac{\partial U_{eff}}{\partial \hat
\beta^{i}}\right|_{\hat \beta =0} =0 \Longrightarrow \hat R_{i} =
\frac{d_i}{D_0-2}\left( -\sum_{j=1}^n \hat R_{j} +2 U(\phi)
\right)\, . \ee
%%%%%%
{}From its structure (a constant on the l.h.s. and a dynamical
function of $\phi (x)$ on the r.h.s) it follows that a
stabilization of the internal space scale factors can only occur
when the nonlinearity field $\phi (x)$ is stabilized as well. In
our freezing scenario this will require a minimum with respect to
$\phi$:
%%%%%%%
\be{1.33} \left. \frac{\partial U(\phi )}{\partial \phi
}\right|_{\phi_{0}} = 0\; \Longleftrightarrow\;  \left.
\frac{\partial U_{eff}}{\partial \phi}\right|_{\phi_{0}}  = 0\, .
\ee
%%%%%%%
We arrived at a stabilization problem, some of whose general
aspects have been analyzed already in Refs.
\cite{GZ(PRD1997)}-\cite{GZ(PRD2000)} and
\cite{GMZ(PRDa)}-\cite{GZBR}. For brevity we only summarize the
corresponding essentials as they will be needed for more detailed
discussions in the next sections.
\begin{enumerate}
\item \label{c1} Eq. \rf{1.32} implies that the scalar curvatures
$\hat R_i$ and with them the compactification scales
$e^{\beta^i_{0}}$ [see relation \rf{1.31a}] of the internal space
components are finely tuned
%%%%%%%%%%%%
\be{1.34} \frac{\hat R_i}{d_i} = \frac{\hat R_j}{d_j}\, , \quad
i,j = 1,\ldots ,n \, . \ee
%%%%%%%%%%%%
\item \label{c2} The masses of the normal mode excitations of the
internal space scale factors (gravitational excitons/radions) and
of the nonlinearity field $\phi$ near the minimum position are
given as \cite{GZ(PRD2000)}:
%%%%%%%%%%%
\ba{1.35} m_{1}^2 &=& \dots \; = m_{n}^2 = \, -\frac{4}{D-2}
U(\phi_{0})\,= -2\frac{\hat R_{i}}
{d_i} > 0\, , \label{1.35a} \\
&\phantom{-} & \nn\\
m_{\phi }^2 &:=& \left. \frac{d^2 U(\phi )}{d \phi^2}
\right|_{\phi_{0}}>0\, .\label{1.35b} \ea
%%%%%%%%%%%%%%
\item \label{c3} The value of the effective potential at the
minimum plays the role of an effective 4D cosmological constant of
the external (our) spacetime $M_0$:
%%%%%%%%
\be{1.36} \Lambda_{{eff}} :=\left. U_{eff}\vphantom{\int} \right|_
{\genfrac{}{}{0pt}{1}{\hat \beta^i =0,}{\phi = \phi_{0}}}\; =\,
\frac{D_0-2}{D-2} U(\phi_{0})\, =\, \frac{D_0-2}{2}\frac{\hat
R_{i}}{d_i}\, . \ee
%%%%%%%%
\item \label{c4} Relation \rf{1.36} implies \be{1.36c4} \sign
\Lambda_{eff} =\sign U(\phi_{0}) =\sign R_{i}\, . \ee Together
with condition \rf{1.35} this shows that in a pure geometrical
model stable configurations can only exist for internal spaces
with negative curvature\footnote{Negative constant curvature
spaces $M_i$ are compact if they have a quotient structure: $M_i =
H^{d_i}/\Gamma_i$, where $H^{d_i}$ and $\Gamma_i$ are hyperbolic
spaces and their discrete isometry group, respectively.}:
%%%%%
\be{1.36c4-4}R_{i} <0\qquad (i=1,\ldots ,n)\, . \ee
%%%%%
Additionally, the effective cosmological constant $\Lambda_{eff}$
as well as the minimum of the potential $U(\phi )$ should be
negative too:
%%%%%%
\be{1.36c4a} \Lambda_{eff} <0, \qquad U(\phi_{0})<0\, . \ee
%%%%%
\end{enumerate}
Plugging the potential $U(\phi)$ from Eq. \rf{1.8} into the
minimum conditions \rf{1.33}, \rf{1.35b}  yields with the help of
$\partial_\phi \bar R=A f'/f''$ the conditions
%%%%%%%
\ba{1.11}
\left.\frac{d U}{d\phi}\right|_{\phi_{0}} &=&
\left.\frac{A}{2(D-2)}\left(f'\right)^{-D/(D-2)}h\right|_{\phi_{0}}
= 0, \quad h:=Df-2\bar R f', \quad \Longrightarrow \quad \
h(\phi_0) = 0\, ,\label{1.11a}\\
\left.\frac{d^2 U}{d\phi^2}\right|_{\phi_{0}} &=& \frac12 A
e^{(A-B)\phi_{0}} \left[ \partial_\phi \bar R + (A-B)\bar R
\right]_{\phi_{0}}= \left.\frac{1}{2(D-1)}
 \left(f'\right)^{-2/(D-2)} \frac{1}{f''}\partial_{\bar R}
 h \right|_{\phi_{0}}>0\, ,\label{1.11b}
\ea
%%%%%
where the last inequality can be reshaped into the suitable form
%%%%%%
\be{1.11c} \left.f''\partial_{\bar R}
h\right|_{\phi_{0}}=f''\left[(D-2)f'-2\bar R f''\right]_{\phi_0}>0
\; . \ee
%%%%%
Furthermore, we find from Eq. \rf{1.11a}
%%%%
\be{1.12-1}
U(\phi_0)=\frac{D-2}{2D}\left(f'\right)^{-\frac{2}{D-2}}\bar
R(\phi_0) \ee
%%%%%%
so that \rf{1.36c4a} leads to the additional restriction
%%%%%
\be{1.13-1}
\bar R(\phi_0)<0
\ee
%%%%%
at the extremum.

In the next section we will analyze the internal space
stabilization conditions \rf{1.34} - \rf{1.36c4a} and \rf{1.11} -
\rf{1.13-1} on their compatibility with particular scalar
curvature nonlinearity $f(\bar R)$. According to our definition
\rf{1.7}, we shall consider the positive branch
%\footnote{The
%$f'<0$ branches were discussed in \cite{Maeda,GMZ(PRDb),GZBR}.}:
%%%%%%
\be{1.43} f'(\bar{R})>0\, .
\ee
%%%%%%
Although the negative $f'<0$ branch can be considered as well (see
e.g. Refs. \cite{Maeda,GMZ(PRDb),GZBR}), we postpone this case for
our future investigations.

%%%%%%%%%%%%%%%%%%%%%%%%%%%%%%%%%%%%%%%%%%%%%%%%%%%%%%%%%%%%%%%%%%%%%%%
\section{The $R^2+R^4$-model\label{model}}
\setcounter{equation}{0}

In this section we analyze a model with curvature-quadratic and
curvature-quartic correction terms of the type
%%%%%%
\be{4.1}
f(\bar{R})=\bar{R}+\alpha\bar{R}^{2}+\gamma\bar{R}^{4}-2\Lambda_{D}\,
. \ee
%%%%%%
We start our investigation for an arbitrary number of dimensions
$D$. First of all, we should define the relation between the
scalar curvature $\bar{R}$ and the nonlinearity field $\phi$.
According to eq. \rf{1.7} we have:
%%%%%%
\be{4.2} f' = e^{A\phi} = 1 +2\alpha \bar{R} + 4\gamma
\bar{R}^3\; . \ee
This equation can be rewritten in the form
%%%%%
\be{4.3}
\bar{R}^{3}+\frac{\alpha}{2
\gamma}\bar{R}-\frac{X(\phi )}{4 \gamma}=0\, ,
\ee
where
%%%%
\be{4.4}
X \equiv e^{A\phi} - 1\, ,\quad -\infty <\phi
<+\infty\:\Longleftrightarrow-1<X<+\infty\, .
\ee
%%%%%
Eq. \rf{4.3} has three solutions $\bar R_{1,2,3}$, where one or
three of them are real valued. Let
%%%%%
\be{s4} q:=\frac{\alpha}{6\gamma},\quad r:=\frac
1{8\gamma}X. \ee
%%%%%
The sign of the discriminant
%%%%%
\be{s5} Q:=r^2+q^3
\ee
%%%%%
defines the number of real solutions (see, e.g., Ref.
\cite{abramowitz}):
%%%%%
\ba{s6} Q>0&\qquad
\Longrightarrow \qquad & \Im \bar R_1=0,\quad \Im \bar R_{2,3}\neq 0\nn\\
Q=0&\qquad \Longrightarrow \qquad & \Im \bar R_i=0 \; ,  i
=1,2,3\, ,
\quad \bar R_1=\bar R_2\nn \\
Q<0&\qquad \Longrightarrow \qquad & \Im \bar R_i=0 \; ,  i
=1,2,3\, .
\ea
%%%%%
It is most convenient to consider $\bar R_i=\bar R_i(X)$ as
solution family depending on the two additional parameters
$(\alpha,\gamma)$. Physical scalar curvatures correspond to real
solutions $\bar R_i(X)$.  For $Q>0$ the single real solution $\bar
R_1$ is given as
%%%%%
\be{s7}\bar
R_1=\left[r+Q^{1/2}\right]^{1/3}+\left[r-Q^{1/2}\right]^{1/3}. \ee
%%%%%
The three real solutions $\bar R_{1,2,3}(X)$ for $Q<0$ read
%%%%%
\ba{s8}
\bar R_1&=&s_1+s_2,\nn\\
\bar R_2&=&\frac 12 (-1+i\sqrt 3)s_1+\frac 12 (-1-i\sqrt 3)s_2=
e^{i\frac{2\pi}3}s_1+e^{-i\frac{2\pi}3}s_2,\nn\\
\bar R_3&=&\frac 12 (-1-i\sqrt 3)s_1+\frac 12 (-1+i\sqrt
3)s_2=e^{-i\frac{2\pi}3}s_1+e^{i\frac{2\pi}3}s_2, \ea
%%%%%
where we can fix the Riemann sheet of $Q^{1/2}$ by setting in the
definitions of $s_{1,2}$
%%%%
\be{s9} s_{1,2}:=\left[r\pm
i|Q|^{1/2}\right]^{1/3}.
\ee
%%%%%

In this paper we investigate the case of positive $Q(\phi)$ that
is equivalent to the condition
%%%%%%
\be{s10}
Q(\phi)>0 \quad \Rightarrow \quad \sign \alpha =\sign \gamma\; .
\ee
%%%%%%
The case $\sign \alpha \neq \sign \gamma \:$ that corresponds to
different signatures of the discriminant $Q$ will be considered in
our forthcoming paper.

To define the conditions for minima  of the effective potential
$U_{eff}$, first we obtain the extremum positions of the potential
$U(\phi )$. The extremum condition \rf{1.11a} for our particular
model \rf{4.1} reads:
%%%%%
\begin{equation}\label{extR}
\bar R_{(0)1}^{4}\gamma\left(\frac{D}{2}-4\right)+\bar
R_{(0)1}^{2}\alpha\left(\frac{D}{2}-2\right)+\bar
R_{(0)1}\left(\frac{D}{2}-1\right)-D\Lambda_{D} =0\, ,
\end{equation}
%%%%%%
where subscript 1 indicates that we seek the extremum positions
for the solution \rf{s7}. Eq. \rf{extR} clearly shows that $D=8$
is the critical dimension for the model \rf{4.1} in full agreement
with the result of the Appendix (see \rf{a8-1}). In what follows,
we investigate this critical case. For $D=8$ eq. \rf{extR} is
reduced to a quadratic one
%%%%%
\begin{equation}\label{extR8}
\bar R_{(0)1}^{2}+\frac{3}{2\alpha}\bar
R_{(0)1}-\frac{4\Lambda_8}{\alpha}=0\;;\quad\Lambda_8\equiv\Lambda_{D=8}
\end{equation}
%%%%%
with the following two roots:
%%%%%
\begin{equation}\label{R0}
\bar
R_{(0)1}^{(\pm)}=-\frac{3}{4\alpha}\pm\sqrt{\left(\frac{3}
{4\alpha}\right)^{2}+\frac{4}{\alpha}\Lambda_8}\quad .
\end{equation}
%%%%%
These roots are real if parameters $\alpha$ and $\Lambda_8$
satisfy the following condition:
%%%%
\begin{equation}\label{con}
\left(\frac{3}{4\alpha}\right)^{2}+\frac{4}{\alpha}\Lambda_8\geq0\;
.
\end{equation}
%%%%%
If $\sign(\alpha)=\sign(\Lambda_8)$, then condition (\ref{con}) is
automatically executed, else
%%%%%
\begin{equation}\label{LL}
|\Lambda_8|\leq\frac{9}{64|\alpha|}\; , \quad \sign(\alpha)\ne
\sign(\Lambda_8)\; .
\end{equation}
%%%%%

To insure that roots (\ref{R0}) correspond to a minimum value of
$U(\phi )$, they should satisfy the condition (\ref{1.11c}):
%%%%%
\be{usmin} f''\left[(D-2)f'-2\bar R f''\right]_{\phi_0}>0 \:
\Longleftrightarrow f''[3+4\alpha \bar R ]_{\phi_0}>0\;, \ee where
\be{} f''=2\alpha+12\gamma\bar R^2\;. \ee
%%%%%

Because for $Q>0$ eq. \rf{s7} is the single real solution of the
cubic eq. \rf{4.3}, then $\bar{R} = \bar{R}_1(\phi )$ is a
monotonic function of $\phi $
%(here, there is one-one mapping between $\phi$ and $\bar{R}_1$).
Thus, the derivative
$\partial_{\phi}\bar R_1=Af'/f''\;$ does not change its sign.
Keeping in mind that we consider the $f'>0$ branch, the function
$\bar{R}_1(\phi)$ is a monotone increasing one for $f''>0$. As
apparent form eq. \rf{s7}, for increasing $\bar{R}_1$ we should
take $\gamma >0$. In a similar manner, the function
$\bar{R}_1(\phi)$ is a monotone decreasing one for $f''
,\gamma<0$.
%and correspondingly for $\gamma<0$.
Thus, for the minimum position $\bar{R}_{(0)1}$, inequality
(\ref{usmin}) leads to the following conditions (we remind that
according to eq. \rf{1.13-1} the minimum position $\bar{R}_{(0)1}$
should be negative and according to eq. \rf{s10} $\sign \alpha =
\sign \gamma $):

\textbf{I.} $\;\; f'' ,\gamma, \alpha >0\;:$
%%%%%%
\be{g>0}3+4\alpha\bar
R^{\pm}_{(0)1}>0 \Longleftrightarrow
|R^{\pm}_{(0)1}|<\frac{3}{4\alpha }\;. \ee
%%%%%%

\textbf{II.} $\;\; f'' ,\gamma , \alpha <0\;:$
%%%%%%%
\be{g<0}3+4\alpha\bar R^{\pm}_{(0)1}<0 \Longleftrightarrow
-|R^{\pm}_{(0)1}|>\frac{3}{4|\alpha|}\;. \ee
%%%%%%%
Obviously, inequality \rf{g<0} is impossible and we arrive to the
conclusion that the minimum of the effective potential $U_{eff}$
is absent if $\sign \alpha = \sign \gamma = -1$.

Additionally, it can be easily seen that in the case
%%%%%
\be{sign}
\sign \alpha = \sign \gamma = \sign \Lambda_D = +1
\ee
%%%%%
the effective potential $U_{eff}$ has no minima also. This
statement follows from the form of the potential $U(\phi )$ for
the model \rf{4.1}. According to eq. \rf{1.8a}, $U(\phi )$ reads:
%%%%%%%
\be{4.8b}
U(\phi ) = (1/2)e^{-B\phi
}\left(\alpha\bar{R}^2+3\gamma\bar{R}^4+2\Lambda_D\right)\, .
\ee
%%%%%%%
Thus, this potential is always positive for parameters satisfying
\rf{sign} and we arrive to the contradiction with the minimum
condition \rf{1.36c4a}. Therefore, the investigation carried above
indicates that the internal space stable compactification is
possible only if the parameters satisfy the following sign
relation:
%%%%%%%%
\be{signs}
\alpha >0, \gamma>0, \Lambda_8<0\, .
\ee
%%%%%%
Let us investigate this case in more detail.
%\be{minsign1}
%\sign \alpha = \sign \gamma = - \sign \Lambda_D
%\ee
%%%%%%
%and
%%%%%%%
%\be{minsign2}
%\sign \alpha = \sign \gamma = \sign \Lambda_D = -1\, .
%\ee
%%%%%%%
%
%Therefore, there are only three possible cases which we
%investigate below in sections 4.1 - 4.3.
%
%
%%%%%%%%%%%%%%%%%%%%%%%%%%%%%%%%%%%%%%%%%%%%%%%%%%%%%
%
%\subsection{$\alpha >0, \gamma>0, \Lambda_8<0$}
%
For this choice of signs of the parameters, it can be easily seen
that both extremum values $\bar R^{(\pm)}_{(0)1}$ from eq. \rf{R0}
satisfy the condition \rf{1.13-1}:  $\bar R^{(\pm)}_{(0)1}<0$.
However, the expression
%%%%%%
\begin{equation}\label{n88}
f'\left(\bar R^{(\pm)}_{(0)1}\right) = 1 + 2\alpha
\bar{R}^{(\pm)}_{(0)1} + 4\gamma \bar
{R}^{(\pm)3}_{(0)1} =-\frac{1}{2} \pm \sqrt{\frac{9}{4}-
16\alpha|\Lambda_8|}-4\gamma\left|\bar R^{(\pm)}_{(0)1}\right|^{3}
\end{equation}
%%%%%%
shows that only $\bar R^{(+)}_{(0)1}$ can belong to $f'>0$ branch.
To make $f'\left(\bar R^{(+)}_{(0)1}\right)$ positive, parameter
$\gamma$ should satisfy the condition
%%%%%%
\begin{equation}\label{Lcond}
  \gamma<\frac{-\frac{1}{2}+\sqrt{\frac{9}{4}-
16\alpha|\Lambda_8|}}{4\left|\bar R^{(+)}_{(0)1}\right|^{3}}\quad.
\end{equation}
%%%%%
As apparent from this equation, parameter $\gamma $ remains
positive if $\Lambda_8$ belongs to the interval
%%%%%%
\be{lambda}
\Lambda_8\in\left(-\frac{1}{8 \alpha},0\right)\, .
\ee
%%%%%%
For this values of $\Lambda_8$, the condition \rf{LL} is
automatically satisfied. We also note, that for positive $\alpha$
and negative $\bar R^{(+)}_{(0)1}$ the condition \rf{g>0} is also
satisfied. Taking into account the interval \rf{lambda}, the
corresponding allowed interval for $\gamma $
reads\footnote{Similar interval for the allowed values of $\gamma$
was also found in \cite{GZBR} for the curvature-quartic model.}
%%%%%
\be{gm}
\gamma\in\left(0,\frac{1}{4\left|\bar
R^{(+)}_{(0)1}\right|^{3}}\right)\;.
\ee
%%%%%

Thus, for any positive value of $\alpha$, Eqs. \rf{lambda} and
\rf{gm} define allowed intervals for parameters  $\Lambda_8$ and
$\gamma$ which ensure the existence of a global minimum of the
effective potential $U_{eff}$. Here, we arrive to the required
stable compactification of the internal space. The position of the
minimum $(\beta^1_0,\phi_0 )$ and its value can be easily found
(via the root $\bar{R}^{(+)}_{(0)1}$) with the help of Eq.
\rf{4.2} and corresponding Eqs. from section 3. The Fig.1 - Fig.2
(see Appendix B )
%\rf{fig221}- Fig. \rf{fig223}
demonstrate such minimum for a particular choice of the
parameters: $\alpha=1,\gamma=1,\Lambda_8=-0.1$. To conclude this
section, we want to note that limit $\Lambda_8 \rightarrow 0$
corresponds to $\bar{R}^{(+)}_{(0)1} \rightarrow 0$ which results
in the decompactification of the internal space $\beta^1_0
\rightarrow \infty$.
%degeneration of the minimum of $U_{eff}$.

\section{Conclusions\label{conclu}}
\setcounter{equation}{0}

In our paper we analyze the model with curvature-quadratic and
curvature-quartic correction terms of the type \rf{4.1} and show
that the stable compactification of the internal space takes place
for the sign relation \rf{signs}. Moreover, the parameters of the
model should belong to the allowed intervals (regions of
stability) \rf{lambda} and \rf{gm}. The former one can be
rewritten in the form
%%%%%%
\be{con1}
\Lambda_8 =\frac{\xi}{8\alpha}\, , \quad \xi\in\left(-1,0\right)\,
.
\ee
%%%%%%
Thus, for the root $\bar{R}^{(+)}_{(0)1}$ and parameter $\gamma$
we obtain respectively
%%%%%%%
\be{con2}
\bar{R}^{(+)}_{(0)1} = \frac{\eta}{\alpha}\, , \qquad
\eta\equiv\frac{1}{4}\left(-3+\sqrt{9+8\xi}\right)<0
\ee
%%%%%%
and
%%%%
\be{con3}
\gamma=\frac{\zeta\alpha^{3}}{4|\eta|^{3}}\; ,\quad
\zeta\in\left(0,1\right)\; .
\ee
%%%%
Eq. \rf{con2} shows that $\bar{R}^{(+)}_{(0)1} \in
\left(-\frac{1}{2\alpha},0\right)$.

It is of interest to estimate the masses of the gravitational
excitons \rf{1.35a} and of the nonlinearity field $\phi$
\rf{1.35b} as well as the effective cosmological constant
\rf{1.36}. From Eqs. \rf{con1}-\rf{con3} follows that
$\bar{R}^{(+)}_{(0)1} \sim \Lambda_8 \sim
-\alpha^{-1}\:,\;\gamma\sim \alpha^{3} \Longrightarrow f'(\phi_0)
\sim \mathcal{O}(1)\; , f''(\phi_0) \sim \alpha\; ,
 U(\phi_0) \sim -\alpha^{-1}$. Then, the corresponding estimates
 read:
%%%%%%
\be{con4}
-\Lambda_{eff}\sim m^{2}_{1}\sim m^{2}_{\phi}\sim \alpha^{-1}\, .
\ee
%%%%%%%
From other hand (see Eqs. \rf{1.31a} and \rf{1.35a})
%%%%%
\be{con5}
U(\phi_0) \sim \exp (-2\beta^1_0) = b_{(0)1}^{-2}\, .
\ee
%%%%%
So, if the scale factor of the stabilized internal space is of the
order of the Fermi length: $b_{(0)1} \sim L_F \sim 10^{-17}$cm,
then $\alpha \sim L_F^2$ and for the effective cosmological
constant and masses we obtain: $-\Lambda_{eff}\sim m^{2}_{1}\sim
m^{2}_{\phi}\sim 1\mbox{TeV}^2$.

In the present paper the analysis of the internal space stable
compactification was performed in the case $Q(\phi )>0 \Rightarrow
\sign \alpha = \sign \gamma $. In our forthcoming paper we extend
this investigation to the case of negative $Q(\phi )$ where the
function $\bar{R}(\phi )$ has three real-valued branches.
%%%%%%%%%%%%%%%%%%%%%%%%%%%%%%%%%%%%%%%%%%%%%%%%%%%%%%%%%%%%%%%%%%%%%%%%%%%%%

%%%%%%%%%%%%%%%%%%%%%%%%%%%%%%%%%%%%%%%%%%

\mbox{} \\ {\bf Acknowledgments}\\ The authors thank Uwe
G$\ddot{u}$nther for the valuable comments.
%%%%%%%%%%%%%%%%%%%%%%%%%%%%%%%%%%%%%%%%%%%%%%%%%%%%%%%%%%%%%%%%%%%%%%%%%%%%%

%%%%%%%%%%%%%%%%%%%%%%%%%%%%%%%%%%%%%%%%%%%%%%%%%%%%%%%%%%%%%%%%%%%%%%%%%%%%%

\appendix
\section{On critical dimensions in nonlinear models\label{app1}}
\setcounter{equation}{0}

The existence of a critical dimension (in our case $D=8$) is a
rather general feature of gravitational theories with polynomial
scalar curvature terms (see, e.g., Refs. \cite{BC,paul,BMS}).
Following our paper \cite{GZBR}, it can be easily demonstrated for
a model with curvature nonlinearity of the type
%%%%
\be{a6} f(\bar
R)=\sum_{k=0}^N a_k \bar R^k \ee
%%%%
for which the ansatz
%%%%%
\be{a7} e^{A\phi}=f'=\sum_{k=0}^N k a_k \bar
R^{k-1} \ee
%%%%%
leads, similar like \rf{1.8a}, to a potential
%%%%%
\be{a8}
U(\phi)=\frac12 \left(f'\right)^{-D/(D-2)}\sum_{k=0}^N (k-1)a_k
\bar R^k. \ee
%%%%%%
The condition of extremum \rf{1.11a} for this potential reads:
%%%%%
\be{a8-1} Df - 2\bar R f' = \sum_{k=0}^N \left(D-2k\right)a_k \bar
R^k =0\, . \ee
%%%%
Thus, at the critical dimension $D=2N$ the degree of this equation
is reduced from $N$ to $N-1$. In this case the search of extrema
is considerably simplified.

In the limit $\phi\to +\infty$ the curvature will behave like
$\bar R\approx c e^{h \phi}$ where $h$ and $c$ can be defined from
the dominant term in \rf{a7}:
%%%%%
\be{a8-1a}e^{A\phi}\approx N a_N \bar
R^{N-1}\approx N a_N c^{N-1} e^{(N-1)h\phi}. \ee
%%%%%
Here the requirement $f'>0$ allows for the following sign
combinations of the coefficients $a_N$ and the curvature
asymptotics $\bar R (\phi\to \infty)$:
%%%%%
\ba{a8-2} N=2l: & \qquad &
\sign [a_N ] = \sign [\bar R (\phi\to
\infty)]\nn\\
N=2l+1: & & a_N >0, \quad \sign [\bar R (\phi\to \infty)]=\pm 1.
\ea
%%%%%%
The other combinations, $N=2l: \ \ \sign [a_N ] =  -\sign [\bar R
(\phi\to \infty)]$, \  $N=2l+1: \ \ a_N <0, \ \sign [\bar R
(\phi\to \infty)]=\pm 1$, would necessarily correspond to the
$f'<0$ sector, so that the complete consideration should be
performed  in terms of the extended conformal transformation
technique of Ref. \cite{Maeda}. Such a consideration is out of the
scope of the present paper and we restrict our attention to the
cases \rf{a8-2}. The coefficients $h$ and $c$ are then easily
derived as $h=A/(N-1)$ and $c=\sign(a_N)\left|N
a_N\right|^{-\frac{1}{N-1}}$. Plugging this into \rf{a8} one
obtains
%%%%%
\be{a9} U(\phi\to +\infty)\approx
\sign(a_N)\frac{(N-1)}{2N}\left|Na_N\right|^{-\frac{1}{N-1}}
e^{-\frac{D}{D-2}A\phi}e^{\frac{N}{N-1}A\phi} \ee
%%%%%%
and that the exponent \be{a10} \frac{D-2N}{(D-2)(N-1)}A \ee
changes its sign at the critical dimension $D=2N$:
%%%%%%%
\be{a11} U(\phi
\to +\infty)\to
\sign(a_N)\frac{(N-1)}{2N}\left|Na_N\right|^{-\frac{1}{N-1}}\times
\left\{ \begin{array}{c}
  \infty \\
  1 \\
  0
\end{array}\right. \quad \begin{array}{c}
  \mbox{for}\ D>2N\, , \\
  \mbox{for}\ D=2N\, , \\
  \mbox{for}\ D<2N\, .
\end{array}
\ee
%%%%%%

This critical dimension $D=2N$ is independent of the concrete
coefficient $a_N$ and is only defined by the degree $\deg_{\bar R}
(f)$ of the scalar curvature polynomial $f$. {}From the
asymptotics \rf{a11} we read off that in the high curvature limit
$\phi\to +\infty$, within our oversimplified classical framework,
the potential $U(\phi)$ of the considered toy-model shows
asymptotical freedom for subcritical dimensions $D<2N$, a stable
behavior for $a_N>0$, $D>2N$ and a catastrophic instability for
$a_N<0$, $D>2N$. We note that this general behavior suggests a way
how to cure a pathological (catastrophic) behavior of polynomial
$\bar R^{N_1}-$nonlinear theories in a fixed dimension $D>2N_1$:
By including higher order corrections up to order $N_2>D/2$ the
theory gets shifted into the non-pathological sector with
asymptotical freedom. More generally, one is even led to
conjecture that the partially pathological behavior of models in
supercritical dimensions could be an artifact of a polynomial
truncation of an (presently unknown) underlying non-polynomial
$f(\bar R)$ structure at high curvatures
--- which probably will find its resolution in a strong coupling
regime of $M-$theory or in loop quantum gravity.

%%%%%%%%%%%%%%%%%%%%%%%%%%%%%%%%%%%%%%%%%%%%%%%%%%%%%%%%%%%%%%%%%%%%%%

\section{Graphical visualizations \label{app2}}
\setcounter{equation}{0}

Following section \ref{model}, we consider the model with one
internal space and critical dimension $D=8$ (as usual, for the
external spacetime $D_0=4$). Then, the effective potential
\rf{1.31} reads:
%%%%%%
\be{b1}
U_{eff}(\hat{
\beta^1},\phi)=e^{-4\hat{\beta^1}}\left[-\frac{1}{2}
\hat{R}_{1}e^{-2\hat{\beta^1}}+U(\phi)\right]\; .
\ee
%%%%%
To draw this effective potential, we define $\hat{R}_{1}$ via
$U(\phi_0)$ in Eq. \rf{1.35a}. In its turn, $U(\phi_0)$ is defined
in Eq. \rf{1.12-1} where $\bar{R}(\phi_0) =
\bar{R}^{(+)}_{(0)1}$ and $f'\left(\bar R^{(+)}_{(0)1}\right)$ can
be found from \rf{n88}. In Figs. 1,2 the generic form of the
$U_{eff}$ is illustrated by a model with parameters
$\alpha=1,\gamma=1,\Lambda_{8}=-0.1$ from the stability regions
\rf{lambda} and \rf{gm}.

\begin{figure}[htbp]
\centerline{\includegraphics[width=2.5in,height=2.5in]{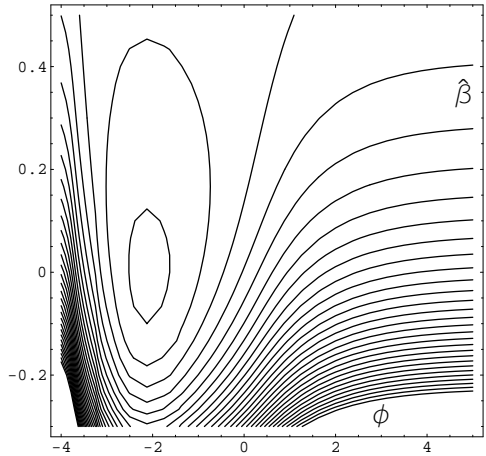}}
\caption{Typical contour plot of the effective potential
$U_{eff}(\hat{
\beta^1},\phi)$ given in Eq. \rf{b1}
with parameters $\alpha=1,\gamma=1,\Lambda_{8}=-0.1$. $U_{eff}$
reaches the global minimum at $(\hat{\beta}^1 =0, \phi\approx
-2.45)$.
\label{fig221}}
\end{figure}

\begin{figure}[htbp]
\centerline{\includegraphics[width=3in,height=3in]{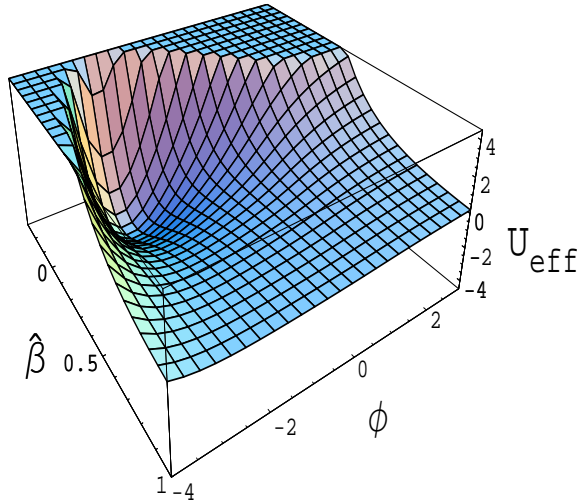}}
\caption{Typical form of the effective potential $U_{eff}(\hat{
\beta^1},\phi)$ given in Eq. \rf{b1}
with parameters $\alpha=1,\gamma=1,\Lambda_{8}=-0.1$.
\label{fig222}}
\end{figure}

%%%%%%%%%%%%%%%%%%%%%%%%%%%%%%%%%%%%%%%%%%%%%%%%%%%%%%%%%%%%%%%%%%%%%%%%%%%%

%%%%%%%%%%%%%%%%%%%%%%%%%%%%%%%%%%%%%%%%%%%%%%%%%%%%%%%%%%%%%%%
%%%%%%%%%%%%%%%%%%%%%%%%%%%%%%%%%%%%%%%%%%%%%%%%%%%%%%%%%%%%%%%

\end{document}